\documentclass[12pt]{book}

\usepackage[dvips]{graphicx,color}
\usepackage{makeidx,observe}
\def\mpc{\,h^{-1}{\rm Mpc}}
\def\kpc{\,h^{-1}{\rm kpc}}
\def\pppm{\rm P^3M}

\def\himsun{{h^{-1}M_\odot}}
\newcommand{\himpc}{{\hbox {$h^{-1}$}{\rm Mpc}} }

\makeauthorindex
\makeindex

\BookTitle{New Trends in Theoretical and Observational Cosmology}
\CopyRight{\copyright 2001 by Universal Academy Press, Inc.}

\begin{document}
\BookTitle{\itshape New Trends in Theoretical and Observational Cosmology}
\CopyRight{\copyright 2001 by Universal Academy Press, Inc.}
\pagenumbering{arabic}

\chapter{
High-resolution numerical simulations for galaxy formation}

\author{%
Y.P. Jing\\
{\it Shanghai Astronomical Observatory, the Partner Group of MPI f\"ur
Astrophysik, Nandan Road 80,  Shanghai 200030, China}
\\
Yasushi Suto \\
{\it Department of Physics and Research Center for the Early Universe
(RESCEU) School of Science, University of Tokyo, Tokyo 113-0033,
Japan}}
\AuthorContents{Y.P. Jing and Y. Suto} 

\AuthorIndex{Jing}{Y.P.}
\AuthorIndex{Suto}{Y.} 

\section*{Abstract}
Understanding of the structure evolution in the Universe has been
greatly advanced with a rapid progress in high-resolution cosmological
simulations. In this contribution, we report a new set of cosmological
simulations with $512^3$ ($\sim 1.4\times 10^8$) particles in a
simulation box of $100\mpc$. With this unprecedented resolution, we have
successfully overcomed the over-merging problem while retaining a
cosmological volume size that is crucial for the statistical analysis.
The simulations are being applied to studying important astrophysical
problems that are sensitive to the simulation resolution.
\section {Introduction}
Galaxy formation is one of the most important and most challenging
problems in astrophysics. In the standard theories of galaxy
formation, it is generally believed that the Universe is dominated by
dark matter (DM), the early density perturbations generated in the
inflationary era are amplified by gravitational instability, the DM
halos form via the collapse of the high density regions, and galaxies
are formed from the cooled gas within those DM halos.

The formation of DM halos, more generally the DM distribution on small
scales, however, is a strongly non-linear problem. Direct numerical
simulations provide the only reliable tool to approach this
problem. Although there have been many exciting developments in
analytical, semi-analytical or empirical modeling of the DM
distribution, these results are generally calibrated with N-body
simulations and their validity still depends heavily on the resolution
of the simulations used.

In this contribution, we report a new set of cosmological N-body
simulations that were obtained recently under a collaboration of
Shanghai Astronomical Observatory and the University of Tokyo. We use
$512^3$ particles to simulate a Cold Dark Matter (CDM) model in a
cosmological volume of $100(\mpc)^3$. The simulations have achieved a
mass resolution which is an order of magnitude better than the advanced
simulations of the Virgo Consortium (Jenkins, A.~et al.\ 1998) as well
as our previous simulations (Jing \& Suto 1998). They are now being
applied to studying a number of astrophysical problems that are
sensitive to the simulation resolutions. As their first application, we
(Jing \& Suto 2002) have studied the internal matter distributions
within virialized halos and have successfully found an ellipsoid
description for the halo density profile that is more accurate than the
conventional spherical description.
\begin{figure}[t]
  \begin{center}
	\vskip 6cm
  \end{center}
  \caption{Dark matter distribution around the most massive halo in the LCDM100a simulation. The slice is $100\mpc$ wide and $10\mpc$ thick.}
\end{figure}

\begin{figure}[t]
  \begin{center} 
	\vskip 6cm
   \end{center} \caption{Dark matter distribution within
    the most massive halo in the LCDM100a simulation. The physical
    size of the figure is 2 times of the halo virial radius}
\end{figure}

\section{N-body simulations}

The simulations are generated with our Particle-Particle-Particle-Mesh
($\pppm$) code on the supercomputer VPP5000 at the National Astronomical
Observatory of Japan. The code adopts the standard $\pppm$ algorithm
(Hockney \& Eastwood 1981; Efstathiou et al. 1985), and is fully
vectorized (Jing \& Suto 1998), and has been recently parallelized. Each
simulation uses $512^3$ simulation particles. Five simulations are
generated for two representative cold dark matter (CDM) models of galaxy
formation. The linear DM distribution in CDM models is completely
specified by the density parameter $\Omega_0$, the cosmological constant
$\lambda_0$, the shape $\Gamma$ and the normalization $\sigma_8$ of the
linear power spectrum.  Table~1 summarizes the physical and simulation
parameters used for these simulations.  The physical parameters are
chosen so that the models are consistent with most observations of the
real Universe.  A box size of $100\mpc$ implies that the particle mass
in this system is only $m_p\approx 10^{9}\himsun$. The force resolution
is $\eta=3\sim 6\kpc$ (the Plummer form), so typical galactic halos can
be well resolved.  Each model (with one box size) will have a few
different realizations in order to properly account for the cosmic
variance (still in progress). One LCDM model uses a small force
softening $\eta=3\kpc$ and is evolved with 5000 time steps, so how the
force softening affects non-linear structures of DM distribution can be
checked. Therefore, these simulations will become an important source
for studying galaxy formation, internal structures of DM halos, and
nonlinear clustering of dark matter.

\begin{table}[h]
\begin{center}
  Table~1.\hspace{4pt} List of simulations \\
\end{center}
\vspace{6pt}
\begin{center}
\begin{tabular}{cccccccccc}
\hline\hline\\[-6pt]
Model & $N$& $\Omega_0$ &  $\lambda_0$ &$\sigma_8$ &$\Gamma$
& $m_p$ ($\himsun$)& $L$ ($\himpc$) & steps & samples\\ 
[4pt]\hline \\[-6pt]
LCDM100a &$512^3$& 0.3  & 0.7 &0.9&0.2 &$6.2\times 10^{8}$&100&5000&1\\
LCDM100 &$512^3$& 0.3  & 0.7 &0.9 & 0.2 &$6.2\times 10^{8}$&100&1200&2\\
SCDM100 &$512^3$& 1.0  & 0.0 &0.55 & 0.5 & $2.1\times 10^{9}$ & 100 &1200 & 2\\
[4pt]\hline 
\end{tabular}
\end{center}
\end{table}

\section{Preliminary results and future work}

Figure 1 shows a dark matter distribution of the LCDM100a simulation
centered at the most massive halo. The plot corresponds to a slice of
$100\mpc$ wide and $10\mpc$ thick. Filamentary structures are very
visible, as many previous studies of such simulations have
discovered. More interesting features of the simulations are seen in
Figure 2 which shows the DM distribution of the most massive
halo. Substructures within the virialized region are preserved well,
similar to high-resolution simulations for individual halos (Moore et
al. 1999), therefore the over-merging problem encountered in previous
studies of cosmological N-body simulations has now been successfully
resolved in our simulations. This will be very important to studying the
dynamical properties and spatial distribution of galaxies when using the
simulations (e.g. based on the semi-analytical modeling).

With this large set of high-resolution simulations, many interesting
cosmological problems will be investigated. As an ``incomplete'' list,
we are 1) studying the DM distribution in the strongly clustering
regimes; 2) analyzing internal structures of DM halos; 3) investigating
galaxy formation by hand-inputing gas physics; 4) applying the
simulations to real observations.

\section*{Acknowledgements} 
We thank Kohji Yoshikawa for invaluable advice on parallelizing the
code.  The work is supported in part by the One-Hundred-Talent Program,
by NKBRSF (G19990754) and by NSFC (No.10043004). The Numerical
simulations were carried out at ADAC (the Astronomical Data Analysis
Center) of the National Astronomical Observatory, Japan.


\begin{thebibliography}{99}  
\bibitem{ApJS}Efstathiou, G., Davis, M., Frenk, C.S. \& White, S.D.M. 1985,
ApJS, 57, 241
\bibitem{}Hockney, R.W.  \& Eastwood, J.W. 1981, Computer simulations using particles. Mc Graw-Hill
\bibitem{ApJ} Jenkins, A.~et al.\ 1998, ApJ, 499, 20
\bibitem{ApJ} Jing, Y.P., Suto, Y., 1998, ApJ, 494, L5
\bibitem{ApJ} Jing, Y.P., Suto, Y., 2002, ApJ, to be submitted
\bibitem{ApJ} Moore, B., Ghigna, S., Governato, F., Lake, G.,
Quinn, T., Stadel, J., \& Tozzi, P., 1999, ApJ, 524, L19
\end{thebibliography}
\end{document}